%% file: main.tex
\documentclass[a4paper,twocolumn,11pt, accepted=2026-01-07]{quantumarticle}
\pdfoutput=1
\makeatletter
\newcommand*{\centerfloat}{%
  \parindent \z@
  \leftskip \z@ \@plus 1fil \@minus \textwidth
  \rightskip\leftskip
  \parfillskip \z@skip}
\makeatother
\usepackage{subcaption}
\usepackage{physics}
\usepackage[T1]{fontenc}
\usepackage{todonotes}
\usepackage{parskip}
\usepackage{amsmath}
\usepackage{amsfonts,amssymb,amsthm,bbm}
\usepackage{hyperref}
\usepackage[capitalize]{cleveref}
\usepackage{mathtools}
\usepackage{xcolor}
\usepackage{svg}
\usepackage{tikz}
\usepackage[labelformat=simple]{subcaption}
\hypersetup{
    colorlinks,
    linkcolor={red!50!black},
    citecolor={blue!50!black},
    urlcolor={blue!80!black}
}
\usepackage{soul}
\usepackage{graphicx}   
\usepackage{float}
\usepackage{amsbsy} 
\usepackage[bold]{hhtensor}
\usepackage{multirow}
\usepackage[deletedmarkup=sout, addedmarkup=bf]{changes}

\definechangesauthor[name={Author}, color=red]{del}
\setdeletedmarkup{\textcolor{red}{\sout{#1}}}

\begin{document}

\title{Minimising the number of edges in LC-equivalent graph states}

\author{Hemant Sharma}
\email{h.sharma-1@tudelft.nl}
\affiliation{QuTech, Delft University of Technology, 2628 CJ, Delft, The Netherlands}
\affiliation{QuSoft and CWI, Science Park 123, 1098 XG Amsterdam, The Netherlands}
\author{Kenneth Goodenough}
\affiliation{College of Information and Computer Sciences,
University of Massachusetts Amherst, Amherst, Massachusetts 01003, USA}
\author{Johannes Borregaard}
\affiliation{Department of Physics, Harvard University, Cambridge, Massachusetts 02138, USA}
\author{Filip Rozp\k{e}dek}
\affiliation{College of Information and Computer Sciences,
University of Massachusetts Amherst, Amherst, Massachusetts 01003, USA}
\author{Jonas Helsen}
\affiliation{QuSoft and CWI, Science Park 123, 1098 XG Amsterdam, The Netherlands}
%\date{30/05/2025}

\begin{abstract}
Graph states are a powerful class of entangled states with numerous applications in quantum communication and quantum computation. Local Clifford (LC) operations that map one graph state to another can alter the structure of the corresponding graphs, including changing the number of edges. Here, we tackle the associated edge-minimisation problem: finding graphs with the minimum number of edges in the LC-equivalence class of a given graph. Such graphs are called minimum edge representatives (MER) and are crucial for minimising the resources required to create a graph state. We leverage Bouchet's algebraic formulation of LC-equivalence to encode the edge-minimisation problem as an integer linear program (EDM-ILP). We further propose a simulated annealing (EDM-SA) approach guided by the local clustering coefficient for edge minimisation. We identify new MERs for graph states with up to 16 qubits by combining EDM-SA and EDM-ILP. We extend the ILP to weighted-edge minimisation, where each edge has an associated weight, and prove that this problem is NP-complete. Finally, we employ our tools to minimise the resources required to create all-photonic generalised repeater graph states using fusion operations.
\end{abstract}
\maketitle

\section{Introduction}\label{sec:introduction}
Graph states are a powerful yet tractable class of multipartite entangled states that have been extensively studied in quantum science. Beyond their theoretical significance~\cite{hein2004multiparty, hein2006entanglement}, they support a wide range of applications, including measurement-based quantum computation~\cite{raussendorf2001one, raussendorf2003measurement, briegel2009measurement, kaldenbach2023, Krishnan_Vijayan_2024, kaldenbach2025}, all-photonic quantum repeaters~\cite{azuma2015all,pant2017rate,rozpkedek2023all,kaur2024resource,patil2024improved,li2024generalised}, loss-tolerant error-correcting codes~\cite{borregaard2020one,bell2023optimising, patil2023tree}, and quantum metrology~\cite{shettell2020graph}. Significant efforts have been directed toward both their experimental realisation~\cite{schwartz2016deterministic, larsen2019deterministic, asavanant2019generation, istrati2020sequential, thomas2022efficient, roh2023generation, cao2023generation, o2024deterministic, ferreira2024deterministic,thomas2024fusion} and theoretical optimisation, encompassing photonic graph state generation~\cite{li2022photonic, takou2024optimisation, Lee_2023, li2024usingreinforcementlearningguide, bhatti2025distributing}, general graph state construction~\cite{cabello2011optimal}, and state distribution~\cite{sim_ann_approach}.

A graph state $\ket{G}$ is represented by a graph $G$ with vertices that correspond to qubits initialised in the $\ket{+}$ state, and edges that correspond to controlled-Z (CZ) gates applied to pairs of qubits. Graph states can also be defined by Pauli stabilizers, and thus constitute a subset of the stabilizer states. By applying local Clifford (LC) gates, i.e.~single-qubit Clifford unitaries, one can transform a graph state $\ket{G}$ into a different \emph{local Clifford equivalent} (or LC-equivalent) graph state $\ket{H}$ with possibly fewer edges in its graph. It was shown in Ref.~\cite{mvdn_2004} that two graph states are LC-equivalent if and only if their graphs are related by a sequence of graph operations called \emph{local complementations}.

A natural goal is to identify LC-equivalent graph states with the fewest edges—referred to as \emph{minimum edge representatives} (MERs). These serve as canonical representatives of \emph{LC-orbits}, the equivalence classes under local complementation~\cite{adcock2020mapping}. MERs have been studied as a means to reduce the physical resources needed for graph state preparation~\cite{cabello2011optimal}. However, identifying MERs is challenging due to the convoluted structure of LC-orbits, as shown in Ref~\cite{adcock2020mapping}. Existing methods typically rely on brute-force enumeration of the entire orbit via local complementations, which quickly becomes intractable since the number of orbit elements can grow exponentially with the number of qubits. As a result, prior work on MERs has been limited to graphs with up to 12 qubits~\cite{cabello2011optimal, adcock2020mapping}.

In this work, we introduce three algorithms to address the edge-minimisation problem: given a graph $G$, find an LC-equivalent graph $H$ with the minimum number of edges. Our first algorithm, EDM-SA, uses simulated annealing~\cite{sim_annealing} to obtain approximate solutions for graphs with up to 100 qubits. The second algorithm, EDM-ILP, is an integer linear program based on Bouchet’s algebraic characterisation of LC-equivalence~\cite{bouchet1991efficient}. While EDM-ILP guarantees globally optimal solutions, its runtime scales exponentially in the worst case. Combining both approaches, our third method, EDM-SAILP, uses EDM-SA to precondition the input to EDM-ILP, yielding a significant constant-factor speed-up. This hybrid approach enables us to compute exact MERs for graphs with up to 16 qubits. Moreover, we extend EDM-ILP to tackle the weighted-edge minimisation problem: finding LC-equivalent graphs that minimise the total edge weight. We prove this problem is NP-complete via a reduction from the vertex-minor problem~\cite{dahlberg2022complexity}.

To demonstrate the practical utility of our approach, we use EDM-SA to optimise the resources required for the creation of all-photonic \emph{generalised} repeater graph states (gRGS)~\cite{li2024generalised}. These states can be utilised for sharing multiple units of entanglement in a quantum network. Moreover, they can be constructed using single-photon sources and fusion operations~\cite{fusion}. However, the probabilistic nature of fusion and photon loss leads to a high number of required single photons and fusion operations, with the fusion order critically affecting generation efficiency. To reduce resource requirements, we identify a sub-graph of the gRGS and construct its corresponding MER, which contains fewer edges and is therefore more efficient to create. LC gates are applied at the start of the protocol to transform the MER into the desired sub-graph, allowing for an optimised fusion order. Using the tools from Ref.~\cite{Lee_2023}, we quantify the resources needed and determine the optimal fusion sequence. Our method achieves more than an order-of-magnitude reduction in resource usage under realistic high-loss conditions. %Additionally, our algorithms can be used to minimise the number of CZ gates needed for general graph state generation, as outlined in Ref.~\cite{cabello2011optimal}.

In Section~\ref{sec: algorithms}, we provide a description of EDM-SA and EDM-ILP algorithms we use to find MERs. In Section~\ref{sec:edge_minimisation}, we study the performance of our algorithms. Next, in Section~\ref{sec: grgs}, we demonstrate the applications of our framework for edge minimisation. We conclude with an outlook for future work in Section~\ref{sec: outlook}.

\section{Algorithms for edge minimisation}\label{sec: algorithms}

\subsection{Simulated annealing}\label{sec: sa_for_em}
Here we discuss a heuristic approach for edge minimisation based on simulated annealing (SA)~\cite{sim_annealing}. Due to its flexibility, simulated annealing has been widely applied in various domains, including the optimisation of resource states in measurement-based quantum computation~\cite{kaldenbach2023, kaldenbach2025}, and the distribution of graph states in networks~\cite{sim_ann_approach}.

SA is an iterative algorithm that runs for a given number of iterations ($k_\textrm{max}$) while trying to approximate the optimal solution. At each iteration $k$, SA may move from a state $s_k$ to another potential `neighbouring' state $s_\textrm{pot}$ with a certain \emph{acceptance probability}. In our implementation, states correspond to graphs, and the state $s_\textrm{pot}$ is sampled from a set of neighbouring states $S(s_k)$. Here, $S(s_k)$ denotes the subset of graph states that differ from the graph state $s_k$ by one \emph{local complementation}. A local complementation $L_v$ is an operation on a graph $G$ that acts on a vertex $v$ by complementing the edges in the neighbourhood of $v$, as shown in Fig.~\ref{fig:lcexample}. We construct the set $S(s_k)$ based on a metric called the local clustering coefficient~\cite{clustering_coeff}, and the degree $D_v$ of a vertex $v$. Details for finding $S(s_k)$ are discussed in Appendix~\ref{sec: choice_vertex}.

Each state has an \emph{energy} associated with it, which reflects its quality, with lower energy states considered to be better candidate solutions. The energy of a state is defined as the number of edges in the graph. The acceptance probability of SA is controlled using the \emph{temperature} parameter ($T(k)$). Transitions from a low- to a high-energy state are more likely at higher temperatures. Thus, starting from a high temperature and slowly cooling down following a defined \emph{temperature schedule}, SA initially explores a wide range of states and gradually converges to a state close to the ground state (state with minimum energy).

\begin{figure}[h!]
\centerfloat
\input{lcexample.tex}\vspace{1mm}
\caption{Local complementation on the red vertex removes (or adds) an edge between the neighbours of the vertex if they were connected (or disconnected) prior to its application}
\label{fig:lcexample}
\end{figure}

For efficient SA, the temperature schedule, i.e., the temperature $T(k)$ at each iteration $k$, must be properly defined. We consider a logarithmic cooling schedule~\cite{cooling_sche}: at the $k$'th iteration, the temperature is given by $T(k) = T(1)/\log_2(k+1) $, for a given initial temperature $T(1)$. {Logarithmic cooling schedule is slower compared to other cooling schedules. This allows EDM-SA algorithm to explore the states for longer time before converging to a solution}. In principle, the initial temperature and $k_\textrm{max}$ can be tailored for specific problem instances. However, we found $T(1)=50$ and $T(1)=100$ to be good enough in practice to find approximate MERs.

\subsection{Edge minimisation using Bouchet's algorithm}\label{sec: ilp_for_em}
% To solve the edge minimisation problem, one can treat the adjacency matrix $A_G$ of the graph $G$ as an input and minimise the edges in $H$ over all the graphs that are LC-equivalent to $G$. This can be written as a minimisation problem as follows:
We leverage Bouchet's LC-equivalence between graph states to find the exact solutions to the edge-minimisation problem. By definition, it can be written as follows:
\begin{align*}
\min_{H} &\hspace{2em}\sum_{i>j}^n A_{H}[i,j],\\
\text{s.t.}&\hspace{2em} G\equiv_{\mathrm{LC}}H ,
\end{align*}
where $A_G~(A_H)$ represents the adjacency matrix of $G~(H)$. We use Bouchet’s algorithm for LC-equivalence~\cite{bouchet1991efficient} to express LC-equivalence in a form amenable to (integer) linear programming. For completeness, we describe Bouchet's algorithm in Appendix~\ref{sec: bouchet_algo}. The algorithm takes as input the adjacency matrices $A_G$ and $A_{H}$ of $n$-vertex graphs $G$ and $H$ respectively, and is based on the binary/symplectic framework (see for example~\cite{dehaene2003local, dehaene2003clifford} and Appendix \ref{sec: bouchet_algo} for further information) to find the following equation describing LC-equivalence of graphs $G$ and $H$,
\begin{gather}\label{eq:bouchet_lin1}
A_G P A_{H} + A_GQ +A_{H}R + S = 0~\textrm{mod}~2\ ,\\
p_i s_i +r_i q_i = 1\ .\label{eq:bouchet_quad1}
\end{gather}

Here, $p_i, q_i,r_i, s_i$ are binary entries of $n\cross n$ diagonal matrices $P,Q,R,S$, respectively. These matrices $P,Q,R,S$ arise in the binary framework of LC operations, where local Clifford operations can be represented by non-singular $2n \cross 2n$ matrices $M$
\begin{align}
    M = \begin{bmatrix}
        P & Q\\
        R & S
    \end{bmatrix},
\end{align}
such that the condition in Eq.~\eqref{eq:bouchet_quad1} corresponds to the symplecticity. We express the LC-equivalence of $G$ and $H$ by setting Eq.~\eqref{eq:bouchet_lin1} and~\eqref{eq:bouchet_quad1} as constraints. Furthermore, we convert Eq.~\eqref{eq:bouchet_lin1} from a binary constraint to a integer one, by introducing a matrix $B$ of integer variables such that:
\begin{align}\label{eq:bouchet_lin2}
A_G P A_{H} + A_GQ +A_{H}R + S = 2B.
\end{align}
We then linearise the equations by introducing constraints to convert products of integer variables into linear systems. This is done for products of variables originating from the terms {$P A_H$} and $A_H R$ in Eq.~\eqref{eq:bouchet_lin2} and terms $PS$ and $RQ$ in Eq.~\eqref{eq:bouchet_quad1}. The approach outlined above produces the full set of constraints; since these are rather unwieldy, we defer the complete derivation of the EDM-ILP algorithm to Appendix~\ref{sec: ilp_derivation}. {We note that Eq.~\eqref{eq:bouchet_lin1} form $n^2$ equations in terms of $4n$ variables, but with the non-linear constraint in Eq.~\eqref{eq:bouchet_quad1}. Bouchet showed in~\cite{bouchet1991efficient} that one only needs to check a small number of solutions, to certify whether there exists a solution that also satisfies Eq.~\eqref{eq:bouchet_quad1}. In particular, if $U$ forms any basis of the vector space $\mathcal{V}$ of solutions to Eq.~\eqref{eq:bouchet_lin1}, then there exists a solution in $\mathcal{V}$ that satisfies the constraint in~\eqref{eq:bouchet_lin2} if and only if there exists an element in the set $\lbrace{u+u'\mid u,u'\in U\rbrace}$ that satisfies the constraint. Note that this latter set has size at most linear in $n$, while $\mathcal{V}$ can in principle be of size exponential in $n$. Unfortunately, we were not able to exploit this structure for the ILP, which has the potential to reduce the number of constraints and thus the runtime significantly. One obstacle encountered is the fact that bases are not naturally expressed in terms of ILPs, making it hard to exploit the above structure. On the other hand, given access to $U$, it is possible to express the set $\lbrace{u+u'\mid u,u'\in U\rbrace}$ naturally in terms of an ILP.}

The runtime of the EDM-ILP algorithm depends on the number of constraints, which in turn depend on the number of vertices and edges in the input graphs. Therefore, we employ EDM-SA to find the approximate MER of a given graph before inputting it into EDM-ILP. This preprocessing step reduces the number of constraints in the corresponding EDM-ILP thus improving the runtime. We refer to this combined algorithm as EDM-SAILP. We use the MOSEK optimiser API for Python~\cite{mosek} to run EDM-ILP and EDM-SAILP. Since EDM-ILP is an exact optimisation, the results obtained are global minima, i.e.~MERs of their corresponding LC-orbit. In Section~\ref{sec:edge_minimisation}, we benchmark the performance of our algorithms for different graph models.

\begin{figure*}[t]
    \centerfloat
    \begin{minipage}{0.49\textwidth}
        \centerfloat
        \includegraphics[width=0.80\linewidth]{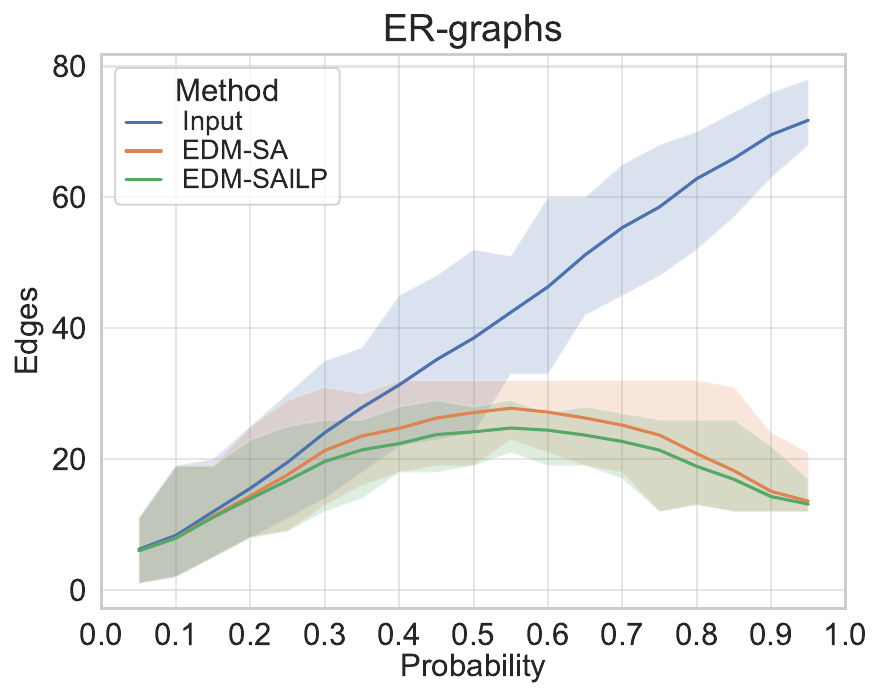}
        \subcaption{}
    \end{minipage}
        \hspace{-5mm}
    \begin{minipage}{0.49\textwidth}
        \centerfloat
        \includegraphics[width=0.80\linewidth]{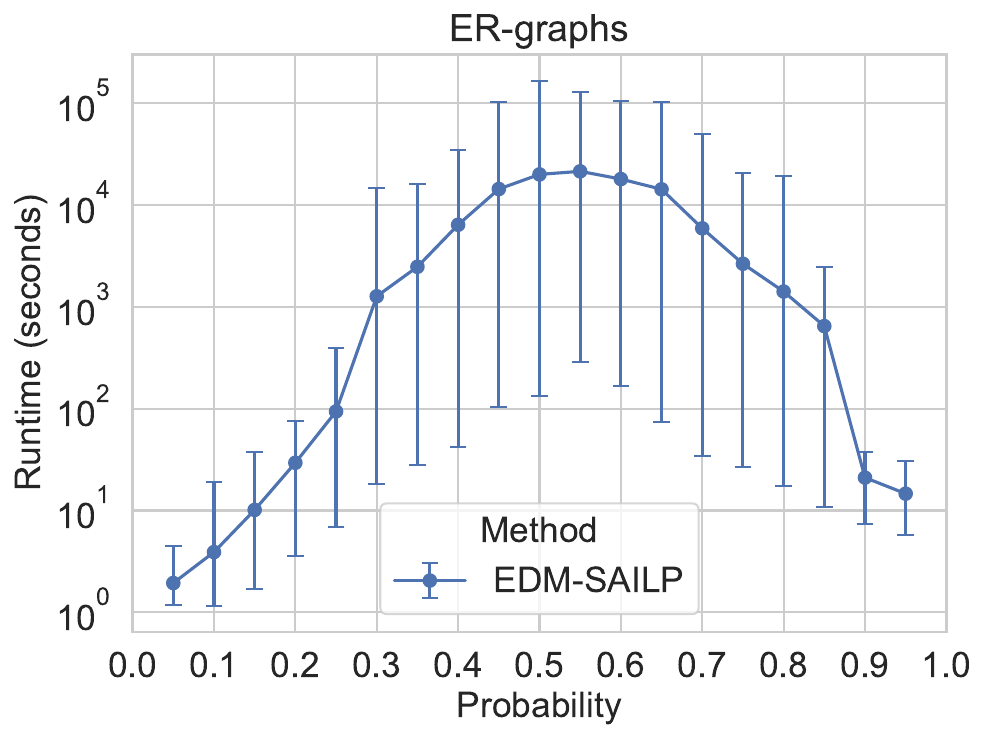}
        \subcaption{}
    \end{minipage}
    \hspace{1mm}
    \begin{minipage}{0.49\textwidth}
        \centerfloat
        \includegraphics[width = 0.80\linewidth]{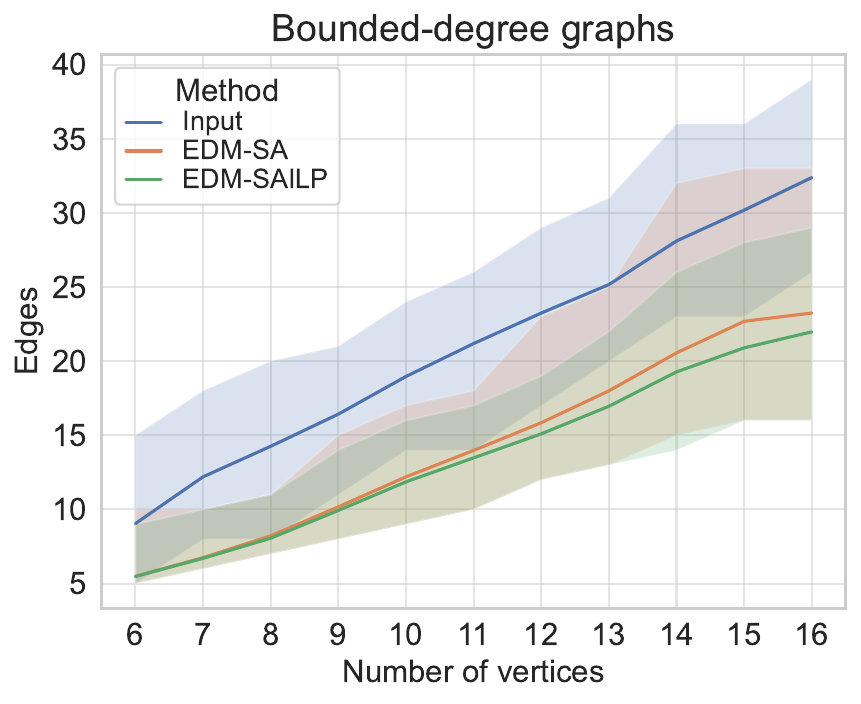}
        \subcaption{}
    \end{minipage}
    \hspace{0mm}
    \begin{minipage}{0.49\textwidth}
        \centerfloat
        \includegraphics[width=0.80\linewidth]{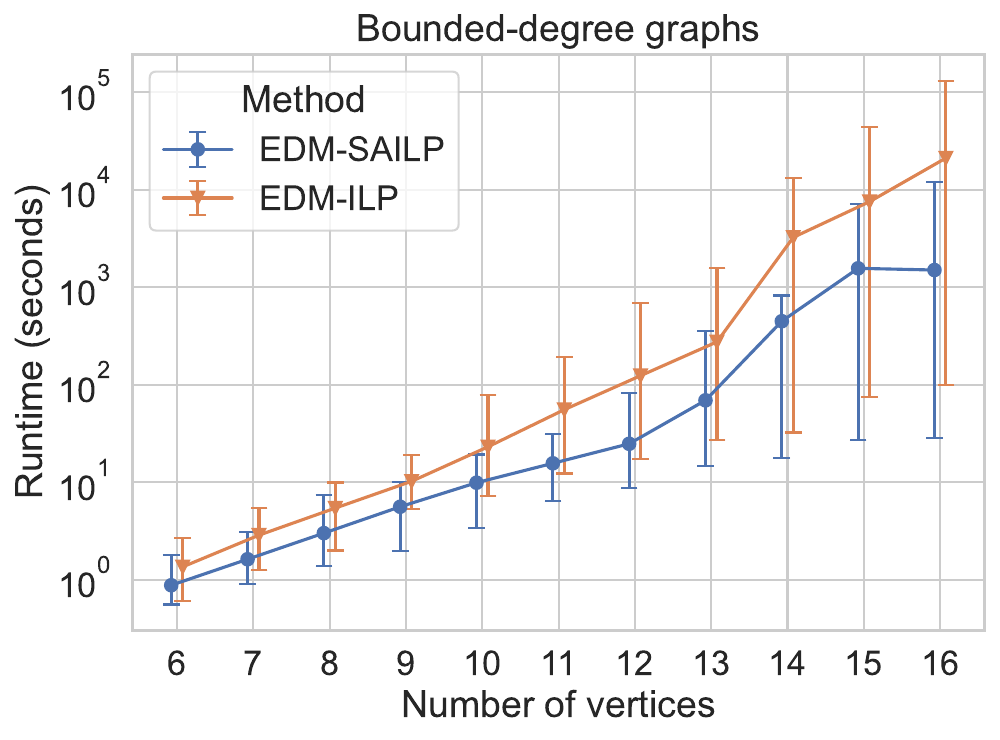}
        \subcaption{}
    \end{minipage}
    \caption{In \textbf{(a)} and \textbf{(c)}, we compare the number of edges in the input graphs and the outputs from EDM-SA and EDM-SAILP for Erd\H{o}s-R\'{e}nyi graphs and bounded-degree graphs respectively. The solid lines denote the average value, and the shaded bands indicate the maximum and minimum. \textbf{(b)} shows the runtime of the EDM-SAILP algorithm as a function of edge inclusion probability for ER graphs with 13 vertices. \textbf{(d)} compares the runtime between EDM-ILP and EDM-SAILP for bounded-degree graphs. The solid line indicates the average runtime, and error bars show the 95th percentile around the mean. For ER graphs, the runtime and the MER edge count are highest around $p\sim 0.5$. From \textbf{(d)}, we observe that EDM-SAILP achieves a constant-factor reduction in runtime compared to EDM-ILP on average. This improvement can be attributed to the EDM-SA preprocessing step, although the runtime remains exponential in the size of the input graph. The average runtime of EDM-SAILP for graphs with 16 vertices is smaller than that for graphs with 15 vertices, which we attribute to a better EDM-SA performance.}
    \label{fig: benchmarking}
\end{figure*}

\subsection{Weighted-edge minimisation}\label{sec:complexity}
%Our proof relies on reducing any instance of the \emph{vertex-minor problem} (see e.g.~\cite{dahlberg2018transforming, dahlberg2020transform, dahlberg2022complexity}) to an instance of a weighted-edge minimisation problem. Since the vertex-minor problem is known to be NP-complete~\cite{dahlberg2020transform, dahlberg2022complexity}, the weighted-edge minimisation problem is NP-complete as well.
Our formulation of Bouchet’s algorithm as an optimisation problem can be easily extended to include weighted-edge minimisation of LC-equivalent graph states. The optimal solution for our case is thus the one that minimises the sum of these weights of LC-equivalent graph states. The optimisation problem then becomes
\begin{align*}
\min_{H} &\hspace{2em}\sum_{i>j}^n A_{H}[i,j]W[i,j],\\
\text{s.t.}&\hspace{2em} G\equiv_{\mathrm{LC}}H \, 
\end{align*}where $W$ is a weighted adjacency matrix encoding the weights $W[i,j]$ of each edge possible pair $i, j$. These weights could represent, for example, the cost of creating edges, thereby enabling a more realistic analysis of the resource requirements for generating graph states, similar to the simulated annealing based approach of Ref.~\cite{sim_ann_approach}. {We allow the weights $W[i,j]$ to be real numbers, since an ILP only requires the variables to be integers, not the cost function.} {Note that the weights $W[i, j]$ are separate from the weights of \emph{weighted graph states}, studied in e.g.~\cite{Anders_2007, Hartmann_2007}, which are not the subject of our work.}

The decision problem for arbitrary weights can be proved to be NP‑complete by reducing the so-called \emph{vertex‑minor problem} to the weighted edge‑minimisation problem. The vertex-minor (or qubit-minor) problem plays an important role in quantum information theory~\cite{dahlberg2018transforming, dahlberg2020transform}. A graph state $\ket{H}$ that can be obtained from $\ket{G}$ under local Clifford unitaries and Pauli measurements is called a qubit-minor of $\ket{G}$~\cite{dahlberg2018transforming, dahlberg2020transform}. Accordingly, the graph $H$ is called a vertex-minor of graph $G$~\cite{oum2005rank}. We consider the labelled version of the vertex-minor problem, where the resultant graph after measurements and local complementations needs to be exactly $H$ (not up to graph isomorphism).  While local equivalence of two graphs $G$ and $H$ can be decided efficiently, deciding whether $H$ is a vertex-minor of $G$ is NP-complete, for both the labelled~\cite{dahlberg2020transform} and unlabelled case~\cite{dahlberg2022complexity}.

The vertex-minor problem can be formulated using the weighted-edge minimisation problem as follows. Assume that the graph $H = (V', E')$ has no isolated vertices~\footnote{For the case of $H$ containing isolated vertices, assign a weight of $0$ to any of the edges incident to an isolated vertex.}. We define the weights $W[i,j]$ according to the following rule:
\begin{gather}
W[i,j] = \begin{cases*}
  -1  & \textrm{if } $\lbrace{i, j\rbrace} \in E' $\\
  +1 & \textrm{if }  $i, j \in V' \land \lbrace{i, j\rbrace} \notin E'$\\
  0 & \textrm{else} 
\end{cases*} \ .
\end{gather}

The smallest weight possible is -$\left|E'\right|$, which is achievable if and only if $G$ is LC-equivalent to a graph $\overline{G}$ such that $\overline{G}\left[V'\right]= H$, i.e.~when $H$ is a vertex-minor of $G$. In other words, an efficient algorithm for deciding whether the minimum score equals $-\left|E'\right|$ would also yield an efficient method for determining whether $H$ is a vertex-minor of $G$, and vice versa. Since the vertex-minor problem is known to be NP-complete in general~\cite{dahlberg2022complexity}, it follows immediately that the decision problem for weighted-edge minimisation is NP-complete as well. We note that this still leaves open whether the minimisation problem remains NP-complete when restricting to uniform/non-negative weights.
%We note that it is not clear if the special case of weighted-edge minimisation when all the weights are unity is NP-complete.

\begin{figure*}[t]
    \centerfloat
    \begin{minipage}{0.99\textwidth}
        \centerfloat
        \includegraphics[width=0.99\textwidth]{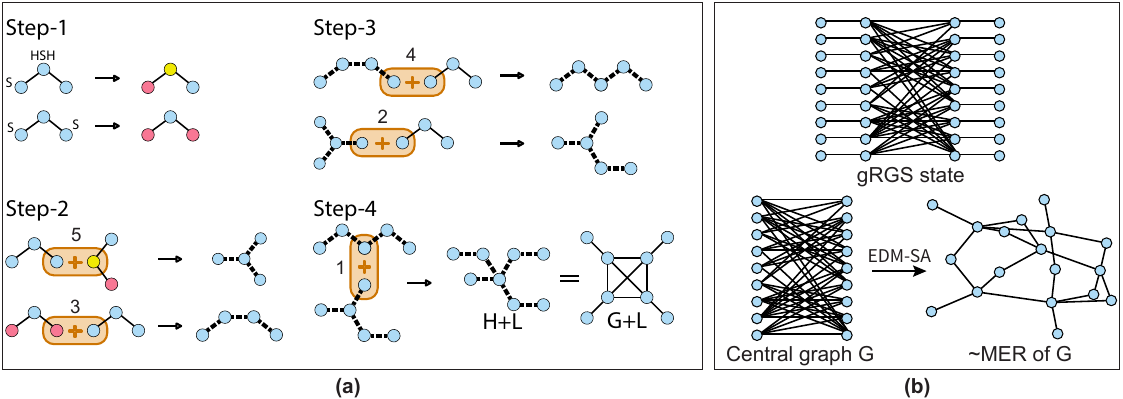}
    \end{minipage}
    \caption{\textbf{(a)} demonstrates the fusion order for creating an RGS with eight qubits, obtained using \texttt{OptGraphState} from Ref.~\cite{Lee_2023}. {The elliptical boxes show the required fusions, labelled by numbers next to them. These numbers allow the tracking of fusions after LC-gates are commuted to the beginning.} The order corresponds to the graph $H$+$L$ with LC-gates applied at the beginning. Graph states (shown with blue vertices and solid edges) may transform into non-graph states when acted upon by LC-gates (represented with red and yellow vertices). The resulting state might also be a non-graph state if such a qubit is involved in fusions, indicated using dashed edges. \textbf{(b)} shows: (1) the sampled gRGS state capable of sharing three ebits across a quantum network, (2) the central graph $G$ (with 46 edges), and (3) its approximate MER $H$, which contains 23 edges.}
    \label{fig: grgs_construction}
\end{figure*}

\section{Edge minimisation of LC-equivalent graph states}\label{sec:edge_minimisation}

In this section, we first characterise the runtime of EDM-SA. We then empirically evaluate: (1) the runtime of EDM-ILP and EDM-SAILP, and (2) the reduction in edge count achieved by EDM-SA and EDM-SAILP. We focus on two graph models: (1) the Erd\H{o}s-R\'{e}nyi model, which is a standard model for creating random graphs, and (2)  the bounded-degree model, which encompasses realistic hardware constraints. Our algorithms ran on each input graph by allocating them on 20 cores of Intel(R) Xeon(R) CPU E5-2620 v2 @ 2.10GHz, each with 32GB of memory. 

We first benchmark our algorithms for the Erd\H{o}s-R\'{e}nyi (ER) graph model~\cite{er_graphs, Gilbert_1959}. An ER graph $G(n,p)$ is a random graph on $n$ vertices, where each of the $\frac{n(n-1)}{2}$ possible edges is included independently with a probability $p$. The expected number of edges in the initial graph increases as the probability $p$ increases. We can thus control the sparsity (or density) of edges in the generated graphs by changing the value of $p$. We first characterise the runtime of the EDM-SA algorithm by running it for different numbers of maximum iterations. The initial temperature is set to $T(1) = 50$ for a set of 100 Erd\H{o}s-R\'{e}nyi graphs with 100 vertices and 2971.2 edges on average. For $k_{\textrm{max}} = 50$, EDM-SA had an average runtime of 0.94 seconds, and the number of edges obtained in the output graphs was 2367.26 on average. Increasing $k_{\textrm{max}}$ to 1050 increases the runtime to 18.82 seconds, and the number of edges obtained was 2310.73. {This suggests that larger values of $k_\textrm{max}$ might allow for better performance of EDM-SA}. However, in the rest of the benchmarking, we set the value of $k_{\textrm{max}}$ and $T(1)$ to 100.

We generate graphs for increasing values of $p$ with $n=13$ vertices using the ER graph model. We generate 100 ER graphs for each value of $p$, and we apply EDM-SA and EDM-SAILP to minimise the edges in each graph generated for a $p$ value. From Fig.~\ref{fig: benchmarking}(a), we find that the number of edges obtained from EDM-SA is on average 1.07 times larger than the MERs for ER graphs. The expected number of edges is concave in $p$, with a maximum at around $p\sim \frac{1}{2}$, and EDM-SA approximations are slightly worse near this regime. Moreover, we observe that dense graphs ($p \geq 0.6$) are LC-equivalent to graphs with relatively fewer edges, shown by the large gap between the input edge count and MER edge count. In Fig.~\ref{fig: benchmarking}(b), the correlation coefficient between the logarithm of runtime of EDM-SAILP and the number of input edges is 0.94, confirming an exponential runtime dependence on the size of input graph. Thus, we omit running the EDM-ILP algorithm for ER graphs in Fig.~\ref{fig: benchmarking}(b) since the number of edges grows to a large number ($\sim 70$) for higher values of probability $p$.

We also benchmark our algorithms using the bounded-degree graph model, which incorporates hardware constraints that caps the maximum degree of edges like restricted qubit connectivity. We generate random graphs with $n$ vertices and bounded degree of $d_{\textrm{lim}}$ by sampling a set of $n$ positive integers where all elements are $\leq d_{\textrm{lim}}$ and constructing simple connected graphs using the Havel-Hakimi algorithm~\cite{Hakimi_1962, Havel1955}. In our simulations, we fixed the value of $d_{\textrm{lim}}$ to 5 and generated graphs with $n$ ranging from 6 to 16. We generated 100 graphs for each vertex number; however for $n=6$ only 63 such graphs were generated.

In Fig.~\ref{fig: benchmarking}(c), we benchmark the edge minimisation performance of EDM-SA against the MERs obtained from EDM-SAILP for bounded-degree graphs. The number of edges obtained from EDM-SA is on average 1.04 times larger than the MERs. We observe that the performance of EDM-SA declines slightly as the number of vertices increases, likely due to the increased sparsity of the input graphs. In addition to edge count comparison, we compare the runtime of EDM-ILP and EDM-SAILP in Fig.~\ref{fig: benchmarking}(d). To find the dependence of runtime of EDM-SAILP, we calculate the correlation coefficient between the logarithm of its runtime and the number of edges and vertices input to the EDM-ILP part of the algorithm. The correlation coefficients between the logarithm of runtime and input edges (vertices) were 0.96 (0.89). Moreover, the addition of EDM-SA as a pre-processing step to EDM-ILP reduces the runtime of the EDM-SAILP algorithm by a factor of two compared to EDM-ILP alone for bounded-degree graphs. Thus, EDM-SAILP is a faster algorithm than EDM-ILP but it still has an exponential runtime dependence on the number of edges and vertices in the input graphs.

\section{Creation of generalised repeater graph states}\label{sec: grgs}

To demonstrate the utility of our algorithms, we consider the generation of photonic repeater graph states for long distance quantum communication. Repeater graph states (RGS)~\cite{azuma2015all} are a key component of all-photonic quantum repeaters, enabling the distribution of a single unit of bipartite entanglement, known as an ebit over long distances. The implementation of such all-photonic repeaters has been extensively studied~\cite{li2022photonic, takou2024optimisation, buterakos2017deterministic, zhan2023performance} but often comes with a large overhead in the number of required single photon sources. A more general class of all-photonic states called \emph{generalised} repeater graph states (gRGS) was introduced in Ref.~\cite{li2024generalised}, which offers a better scaling between photon overhead and the number of shared ebits. Here, we apply our MER algorithms to further minimise the number of single photons and optical \emph{fusion} operations for the creation of gRGS.

\begin{figure*}[t]
    \centerfloat
    \begin{minipage}{0.49\textwidth}
        %\centerfloat
        \includegraphics[width = 0.85\linewidth]{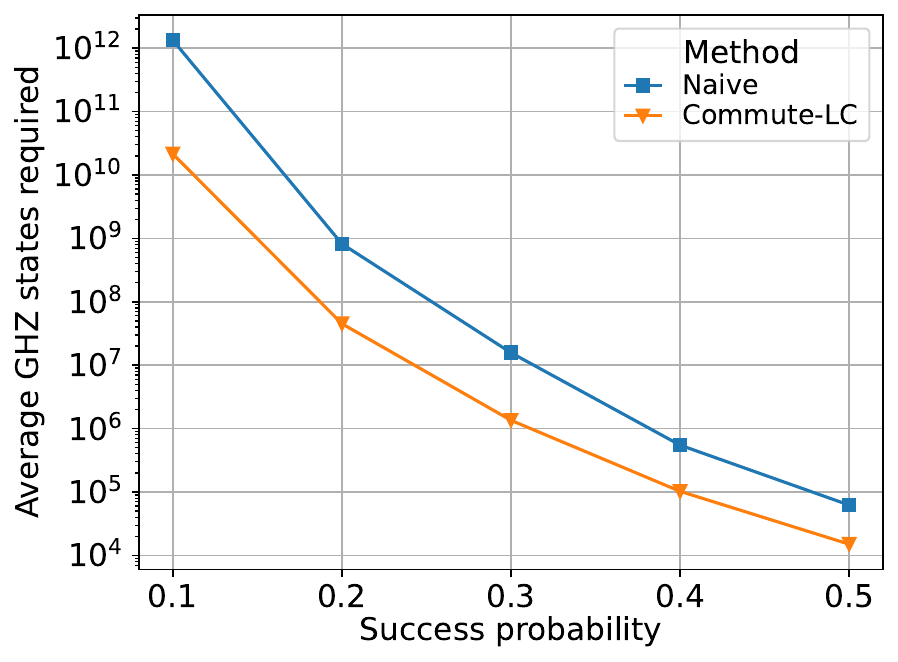}
        \subcaption{}
    \end{minipage}
    \hspace{2mm}
    \begin{minipage}{0.49\textwidth}
        %\centerfloat
        \includegraphics[width=0.85\linewidth]{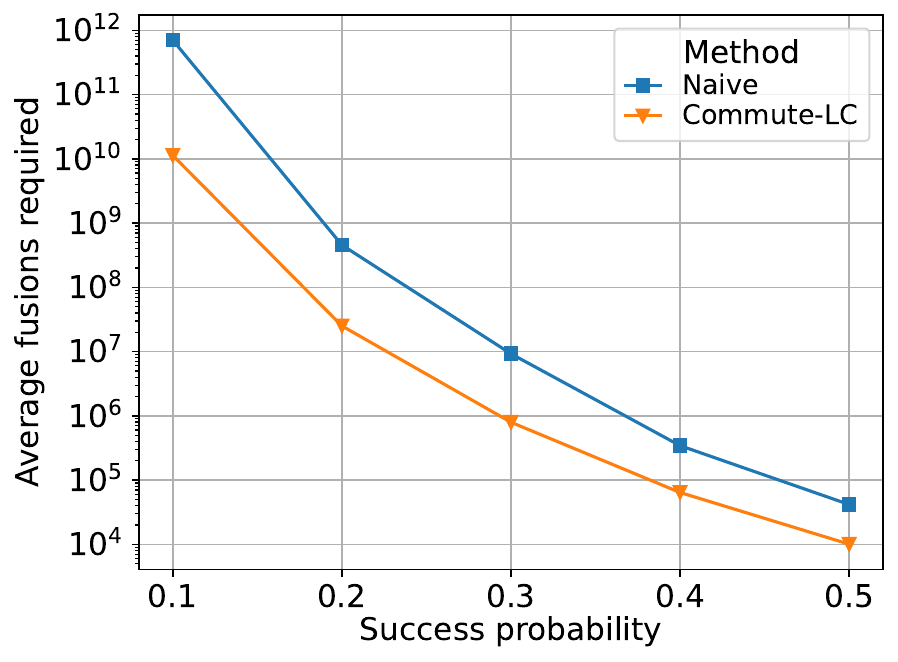}
        \subcaption{}
    \end{minipage}
    \caption{\textbf{(a)} and \textbf{(b)} show the required number of resource states and fusions, respectively, for creating a single instance of $G$+$L$ gRGS with 64 edges and 36 vertices (see Fig.~\ref{fig: grgs_construction}(b)). We compare the naive approach of direct gRGS construction (blue curve) with the Commute-LC protocol (orange curve). The Commute-LC protocol requires nearly two orders of magnitude fewer resources than the naive approach, especially at lower fusion success probabilities. The savings decrease to less than an order of magnitude for relatively higher success probabilities.}
    \label{fig: grgs_perf}
\end{figure*}

Fusions are probabilistic operations that join two graph states into a larger graph state upon success~\cite{fusion}. The success probability of a standard implementation of a fusion is upper-bounded by 50\%. However, in practice the success probability is often lower than 50\% due to hardware losses. The fundamental resource we consider for creating graph states is a 3-qubit GHZ state (resource state). Such a state can be created probabilistically but in a heralded fashion by fusing six single photons~\cite{varnava2008good}.

The probabilistic nature of fusions necessitates a significant amount of resources to construct practically useful graph states. Creating all-photonic graph states involves multiple rounds of fusions starting with resource states that are fused to create intermediate graph states, which are incrementally fused to construct the final graph state. For a given graph, Ref.~\cite{Lee_2023} constructs a so-called \emph{fusion network}, indicating the order in which fusions should occur. In general, it is desirable to fuse two graph states of similar sizes to mitigate the effects caused by fusion failure. Moreover, multiple fusion networks may exist for a given graph, each potentially requiring different amounts of resources. Ref.~\cite{Lee_2023} addresses these challenges and reduces the resource overhead for photonic graph state creation by optimising (1) the number of required fusions, (2) the number of resource states, and (3) the fusion order. Moreover, they provide a numerical tool for this purpose called the \texttt{OptGraphState}.

Here, we exploit the structure of generalised repeater graph states to optimise their experimental creation. Such graphs consist of a densely-connected central graph ($G$) and leaves ($L$) that are attached to each vertex of $G$. We find an approximate MER ($H$) of the central graph $G$ using our EDM-SA algorithm. Naively, one could fuse resource states to create $G$+$L$ directly. However, a more resource-efficient method for creating a gRGS could be: (1) creating the graph $H$ using fusions, (2) applying LC-gates to convert the graph $H$ to $G$, and (3) adding leaves $L$ to $G$. We will refer to this way of construction as \emph{Intermediate-LC}. However, the addition of leaves at the end constrains the order of fusions which could affect the probability of creation of the gRGS. We propose commuting the LC-gates to the beginning of the protocol that lifts this constraint and allows for an optimal fusion order. We refer to this protocol as \emph{Commute-LC}. In conclusion, the Commute-LC protocol has two steps: (1) apply the LC-gates to the resource states at the beginning of the protocol, (2) follow fusion order to create the graph $H$+$L$ using fusions. The LC-gates here correspond to the conversion of the graph $H$ to $G$. We should note that fusions involve the application of single-qubit Pauli correction operations, but they can be commuted through Clifford gates since these are Pauli corrections. 

We illustrate the construction of $G$+$L$ using the Commute-LC protocol in Fig.~\ref{fig: grgs_construction}(a). Following the steps in Commute-LC, we first apply LC-gates to the appropriate qubits, followed by fusing states according to the order obtained for creating $H$+$L$. Applying LC-gates converts the resource states into LC-equivalent \emph{non-graph} states. The Commute-LC protocol fuses these non-graph states following the order of $H$+$L$ to obtain the desired graph state $G$+$L$. Appendix~\ref{sec: grgs_app} details how to determine locations of LC-gates and how following the order for $H$+$L$ constructs the $G$+$L$ graph. We note here that Ref.~\cite{Lee_2023} also applies LC-gates to their qubits during their optimisation. However, these LC-gates only convert a fully-connected sub-graph (clique) into a star. In contrast, by applying EDM-SA to the central graph $G$ in Protocols 1 and 2, we generalise the use of LC-gates to simplify the fusion process not just for cliques but for arbitrary densely connected sub-graphs. 

In Fig.~\ref{fig: grgs_perf}, we compare two methods: (1) the naive approach of creating $G$+$L$ directly, and (2) creation following the Commute-LC protocol. We use the Python package \texttt{OptGraphState} to estimate the required resources. We consider a gRGS that can share 3 ebits across a network~\footnote{Similar to Ref.~\cite{li2024generalised}}, shown in Fig.~\ref{fig: grgs_construction}(b). This gRGS has a central graph ${G}$ with 18 vertices and 46 edges, shown in Fig.~\ref{fig: grgs_construction}(b). We find an approximate MER of this graph with 23 edges using the EDM-SA algorithm (with $k_{\textrm{max}}=10000$ and $T(1)=100$). We observe that commuting LC-gates to the beginning and following fusion order for creating $H$+$L$ significantly reduces the required resources. This reduction is nearly two orders of magnitude for fusions in high-loss conditions.

\section{Outlook}\label{sec: outlook}
We have proposed three algorithms to address the edge-minimisation problem. First, a simulated annealing (EDM-SA) approach efficiently finds approximate MERs for graphs with up to 100 qubits. Second, we introduce an integer linear programming (EDM-ILP) formulation that yields exact solutions, albeit with potentially exponential runtime. Third, we combine both methods into a hybrid EDM-SAILP algorithm, enabling exact MER identification for graphs with up to 16 qubits.

We benchmark all algorithms on two graph models: bounded-degree and Erd\H{o}s-R\'{e}nyi. Additionally, we extend the ILP to handle weighted-edge minimisation and prove that this variant is NP-complete. Importantly, we show that the scalable EDM-SA approach manages to find approximate MERs with less than 10\% more edges on average than the actual MERs for a large set of randomly sampled graphs. {Moreover, the performance of EDM-SA could still be improved by tailoring the initial temperature and maximum iterations for specific problem instances.}

We have applied our EDM-SA algorithm to minimise the required resources  for all-photonic generalised repeater graph states (gRGS) demonstrating its practical utility for long distance quantum communication. Specifically, we have proposed to apply LC-gates corresponding to local complementation to the resource states before fusing them to create the gRGS states. We then demonstrate that EDM-SA can be combined with existing methods from Ref.~\cite{Lee_2023} to reduce the required number of single-photon sources and fusion operations by more than an order of magnitude for realistic lossy implementations. 

%Specifically, we determine fusion orders and the locations of the LC-gates for creating gRGS. 

Our results open up a number of future directions. Finding the MER can reduce the number of required CZ gates for creating a desired graph state between qubits following the idea in Ref.~\cite{cabello2011optimal}. However, there are examples of graph states where the minimum number of edges up to local complementations is greater than the minimum number of two-qubit gates needed to create the state~\cite{kumabe2024complexity,jena2024graph, davies2025preparing}. It would be interesting to further investigate the exact connection between MERs and the minimal number of CZ-gates for generating qubit graph state generation

Furthermore, we have focused exclusively on transformations between graph states that are allowed with local Clifford unitaries. In practical scenarios, however, one might consider arbitrary local unitaries (LU), which define a strictly coarser equivalence relation on graphs than the LC-equivalence relation, as was first shown in~\cite{ji2007lu}. Graph states under local unitary equivalence are significantly less well-understood, but this topic has seen recent investigations~\cite{burchardt2024algorithm, claudet2024local, claudet2025deciding}, closely relating this equivalence to the Clifford hierarchy. One natural follow-up question is whether edge minimisation of graphs under local unitaries can similarly be phrased as an ILP (using, for example, the algorithm from~\cite{claudet2025deciding}), and, if so, to what extent this coarser equivalence can reduce the edge count.

\begin{acknowledgements}
We thank Tim Coopmans and Sebastiaan Brand for useful conversations, and Luise Prielinger for her help with the HPC.
JH acknowledges funding from the Dutch Research Council (NWO) through a Veni grant (grant No.VI.Veni.222.331). JH and HS acknowledge funding from the Quantum Software Consortium (NWO Zwaartekracht Grant No.024.003.037). JB acknowledges support from The AWS Quantum Discovery Fund at the Harvard Quantum Initiative. 
{FR acknowledges support from NSF (NSF grant No. ERC-1941583).}
Part of this work was performed while HS was on a research visit at Harvard University funded by a Quantum Delta NL travel grant.
\end{acknowledgements}

\section*{Code availability}
{The package is publicly available on PyPI and can be installed with \texttt{pip~install~graphstate-opt}.} Moreover, a release of our code is also available at~\cite{_2025_graph_state_optimization} and the accompanying data  at~\cite{sharma_2025_15534839}, with the expectation that it will serve as a valuable resource for the quantum information community in addressing practical optimisation challenges related to graph states. The usage guidelines and notes can be found at the GitHub repository \url{https://github.com/arr0w-hs/graph_state_optimization}.

\bibliographystyle{quantum}
\bibliography{biblio}% Produces the bibliography via BibTeX

\onecolumn\newpage
\appendix

\section{LC-equivalence algorithm for graph states}\label{sec: bouchet_algo}

We will now briefly recap Bouchet's algorithm, using the stabilizer formalism (as was done in~\cite{van2004efficient}). In particular, we will focus on the commonly used symplectic framework of the stabilizer formalism~\cite{dehaene2003local, van2004efficient}. In the symplectic framework, Pauli strings are mapped to elements in $\mathbb{F}_2^{2n}$ through a map $\phi$. Furthermore, conjugation of Pauli strings by Clifford unitaries correspond to symplectic transformations of $\mathbb{F}_2^{2n}$. That is, for each Clifford $C$ there exists a $2n\times 2n$ binary matrix $M$ such that $\phi\left(CsC^\dagger\right)=M\phi\left(s\right)$, for every Pauli string $s$. The matrix $M$ being symplectic means that

\begin{align}
M^T\Omega M = \Omega~\textrm{,  ~where }\Omega = \begin{bmatrix}
    0 & I\\
    I & 0\\
\end{bmatrix}\ , 
\end{align}
and $I$ is the $n\cross n$ identity matrix. This encodes the fact that Clifford transformations need to preserve the commutation relations between Pauli strings. In fact, every matrix that satisfies the above symplectic condition corresponds to some Clifford transformation. In what follows, we will identify Pauli strings and Clifford unitaries and their images under $\phi$.

The generator sets of the two graph states $\ket{G}$ and $\ket{H}$ under consideration are given by

\begin{align}
\begin{bmatrix}
A_G\\
I
\end{bmatrix}, ~\begin{bmatrix}
A_H\\
I
\end{bmatrix}, 
\end{align}

where $I$ is an $n\times n $ identity matrix, and $A_G$ and $A_{H}$ are the adjacency matrices of $G$ and $H$, respectively. Indeed, the $i$'th column of each tableaux corresponds to the $i$'th stabilizer generator. %(see Eq.~\eqref{eq:stab_gens}).

In the symplectic framework, $G$ and $H$ are LC-equivalent if and only if there exist $M$ and $W$ such that 
\begin{align}
&M\begin{bmatrix}
A_H\\
I
\end{bmatrix}W=\begin{bmatrix}
A_G\\
I
\end{bmatrix}\\
 &M = \begin{bmatrix}
        P & Q\\
        R & S
    \end{bmatrix},~ W \in \textrm{GL}_{2n}\left(\mathbb{F}_2\right) \ .
\end{align}

Here $M$ represents single-qubit Clifford gates, and the $P,Q,R,S$ submatrices are $n\cross n$ diagonal matrices with binary entries $p_i, q_i,r_i, s_i$ such that
\begin{equation}\label{eq:bouchet_quad}
p_i s_i +r_i q_i = 1,
\end{equation}
which corresponds to the symplectic constraint. The $W$ matrix encodes the fact that the generating sets need not be equal, but should span the same space. That is, the resulting generator matrices only need to be equivalent up to invertible linear transformations. 

% = \begin{bmatrix}
    %     \textrm{diag}\left(p_1, \ldots, p_n\right) & \textrm{diag}\left(q_1, \ldots, q_n\right)\\
    %     \textrm{diag}\left(r_1, \ldots, r_n\right) & \textrm{diag}\left(s_1, \ldots, s_n\right)
    % \end{bmatrix}

%  A Clifford unitary consisting of single-qubit Clifford gates is represented by a matrix $M$ of the following form
% \begin{align}
%     M = \begin{bmatrix}
%         P & Q\\
%         R & S
%     \end{bmatrix},
% \end{align}
% where, $P,Q,R,S$ are $n\cross n$ diagonal matrices with binary entries $p_i, q_i,r_i, s_i$ such that:
% \begin{equation}\label{eq:bouchet_quad}
% p_i s_i +r_i q_i = 1,
% \end{equation}
% which corresponds to the invertibility of the local Clifford unitaries. 

% The question is now when there exists such an above $M$ such that $M\begin{bmatrix}
% A_H\\
% I
% \end{bmatrix}$ yields a generating set for $\ket{G}$.

As was noted in~\cite{van2004efficient}, one can get rid of the degree of freedom of $W$. This is since the columns of two matrices $K_1, K_2$ of rank $n$ generate the same stabilizer group if and only if $K_1^T\Omega K_1 = K_1^T\Omega K_2 = K_2^T\Omega K_2$ is the zero matrix. It thus suffices to check whether there exists an $M$ such that 
\begin{align}
\begin{bmatrix}A_G &I\end{bmatrix}\Omega M \begin{bmatrix}A_H\\
I\end{bmatrix} = 0\label{eq:convert_condition}\ .
\end{align}

Expanding Eq.~\eqref{eq:convert_condition} yields:
\begin{align}\label{eq:bouchet_lin}
A_G P A_{H} + A_GQ +A_{H}R + S = 0.
\end{align}
A solution vector $(p_i, q_i,r_i, s_i)$ of the above linear system corresponds to an LC operation if and only if it follows the constraint in Eq.~\eqref{eq:bouchet_quad}.

Eq.~\eqref{eq:bouchet_lin} is a set of $n^2$ equations in $4n$ variables $(p_i, q_i,r_i, s_i)$ with a constraint (Eq.~\eqref{eq:bouchet_quad}) that is non-linear, so in principle not easily solvable. However, through an ingenious argument, Bouchet~\cite{bouchet1991efficient} proved that only a small number of solutions to Eq.~\ref{eq:bouchet_lin} have to be checked to verify if there exists one that also satisfies Eq.~\eqref{eq:bouchet_quad},  making the total runtime polynomial~\cite{bouchet1991efficient}. {However for edge minimisation, we have unfortunately not been able to exploit this extra structure with our ILP (see the main text). As such, we just use Eqs.~\ref{eq:bouchet_lin} and \ref{eq:bouchet_quad} to construct our ILPs.}

\section{Derivation of EDM-ILP}\label{sec: ilp_derivation}
In this section, we show how we formulate an ILP for edge minimisation inspired by Bouchet's algorithm. 

Let us first formally state our optimisation problem. The input is a single graph $G$ and the output is an LC-equivalent graph $H$ with a minimal number of edges. Mathematically it is framed as follows:
\begin{align*}
\min_{H} &\hspace{2em}\sum_{i>j}^n A_{H}[i,j],\\
\text{s.t.}&\hspace{2em} G\equiv_{\mathrm{LC}}H.
\end{align*}
% We utilise Bouchet's algorithm as a means to provide (linear and nonlinear) constraints on the (binary matrix) variable $A_{H}$ in terms of additional diagonal matrices $(P,Q,R,S)$ corresponding to local Clifford gates.
We now utilise the formulation of LC-equivalence found in Eqs.~\eqref{eq:bouchet_lin} and~\eqref{eq:bouchet_quad}. Moreover, since the matrices have binary entries we put that as constraints.
This leads to the following equivalent formulation,
\begin{align}\label{eq: ilp1}
\min &\hspace{0.75em}\sum_{i>j}^n A_{H}[i,j],\\
\text{s.t.}&\hspace{0.75em} A_GPA_{H} + A_GQ + A_{H}R + S = 0\hspace{-0.5em}\mod 2,\notag\\
&\hspace{0.75em} A_{H}\in \{0,1\}^{n\times n},\notag\\
&\hspace{0.75em} A_{H}[i,j] = A_H[j,i],\notag\\
&\hspace{0.75em} A_{H}[i,i] = 0,\notag\\
&\hspace{0.75em} P[i,i]S[i,i] + R[i,i]Q[i,i] = 1,\notag\\
&\hspace{0.75em} P[i,j],Q[i,j],R[i,j],S[i,j] = 0, i \neq j,\notag\\
&\hspace{0.75em} P[i,j],Q[i,j],R[i,j],S[i,j] \in \{0,1\} , \notag
\end{align}
where we write the $(i,j)$th element of a matrix $M$ as $M[i,j]$.
While this is an integer program, it is not an integer \emph{linear} program due to the products of variables as well as the modular constraint. To turn this into an ILP we first remove the modulo 2 constraint by introducing a matrix $B\in \mathbb{Z}^{n\times n}$ of integer variables and demanding that
\begin{equation}
A_GP A_{H} + A_GQ + A_{H}R + S =2B\ .\nonumber
\end{equation}
The optimisation problem can thus be written as:
\begin{align}\label{eq: ilp2}
\min &\hspace{0.75em}\sum_{i>j}^n A_{H}[i,j],\\
\text{s.t.}&\hspace{0.75em} A_G[i,a] P[a,b] A_H[b,j] + A_G[i,a] Q[a,j] + \notag\\
&\hspace{1.5em}A_H[i,a] R[a,j] + S[i,j] =2B[i,j]\hspace{-0.0em},\notag\\
&\hspace{0.75em} P[i,i]S[i,i] + R[i,i]Q[i,i] = 1,\notag\\
&\hspace{0.75em} A_{H}[i,j] = A_H[j,i],\notag\\
&\hspace{0.75em} A_{H}[i,i] = 0,\notag\\
&\hspace{0.75em} A_{H}[{i,j}] \in \{0,1\},\notag\\
&\hspace{0.75em} P[i,j],Q[i,j],R[i,j],S[i,j] = 0, i \neq j,\notag\\
&\hspace{0.75em} P[i,j],Q[i,j],R[i,j],S[i,j] \geq 0. \notag
\end{align}
We then turn the quadratic constraints (i.e.~terms corresponding to $A_GPA_H$, $A_HR$, $PS$ and $RQ$) into linear constraints by introducing extra variables. For each term, we first replace each product $xz$ of binary variables $x$ and $z$ with a single binary variable $y$. We then add the following constraints $(1)\, y\leq z,\; (2)\, y\leq x,\;(3)\, y\geq x+z-1$. One thing to keep in mind is that $A_G$ is a constant, so quadratic terms with $A_G$ do not need to be linearised. We replace each product of the form $P[a,b]A_H[b,j]$ and $A_H[i,a]R[a,j]$ with new corresponding variables $Z^P_{abj}$ and $Z^R_{aij}$, respectively. For terms of the form $PS$ and $RQ$ we introduce new variables $Z^{PS}$ and $Z^{RQ}$, respectively. Applying these changes to Eq.~\eqref{eq: ilp2} we obtain our EDM-ILP:
\begin{align}\label{eq: ilp3}
\min &\hspace{0.75em}\sum_{i>j}^n A_{H}[i,j],\\
\text{s.t.}&\hspace{0.75em}A_G[i,a]Z^P_{abj}+A_G[i,a]Q[a,j]+\notag\\
&\hspace{7.25em}Z^R_{aij}+S[i,j]=2B[i,j],\notag \\
&\hspace{0.75em} Z^P_{abj}\leq A_H[b,j],\notag\\ 
&\hspace{0.75em} Z^P_{abj}\leq P[a,b],\notag\\
&\hspace{0.75em} Z^P_{abj}\geq A_H[b,j]+P[a,b]-1,\hspace{0em}\notag\\
&\hspace{0.75em} Z^R_{aij}\leq A_H[i,a],\notag\\
&\hspace{0.75em} Z^R_{aij}\leq R[a,j],\notag\\
&\hspace{0.75em} Z^R_{aij}\geq A_H[i,a]+R[a,j]-1,\notag\\
&\hspace{0.75em} Z^{PS}[i,i] \leq S[i,i],\notag\\
&\hspace{0.75em} Z^{PS}[i,i] \leq P[i,i],\notag\\
&\hspace{0.75em} Z^{PS}[i,i] \geq S[i,i]+P[i,i]-1,\notag\\
&\hspace{0.75em} Z^{RQ}[i,i] \leq R[i,i],\notag\\
&\hspace{0.75em} Z^{RQ}[i,i] \leq Q[i,i],\notag\\
&\hspace{0.75em} Z^{RQ}[i,i] \geq R[i,i]+Q[i,i]-1,\notag\\
&\hspace{0.75em} A_{H}[i,j] = A_H[j,i],\notag\\
&\hspace{0.75em} A_{H}[i,i] = 0,\notag\\
&\hspace{0.75em} A_{H}[{i,j}] \in \{0,1\},\notag\\
&\hspace{0.75em} P[i,j],Q[i,j],R[i,j],S[i,j] = 0, i \neq j,\notag\\
&\hspace{0.75em} P[i,j],Q[i,j],R[i,j],S[i,j] \in \{0,1\}.\notag
\end{align}

Using the ILP, we can thus find LC-equivalent graphs to a given graph that have minimum number of edges. We note here that better way{s} of solving the edge-minimisation problem using ILP might exist that could allow us to go beyond graph states with 16 qubits.

\section{Choice of neighbouring states in EDM-SA}\label{sec: choice_vertex}
In our implementation of SA, the state $s_k$ at the $k$'th iteration is given by a graph. A natural choice for the set $S(s_k)$ of neighbouring states is the set of graphs that differ from $s_k$ by a single local complementation. This set could have a maximum size equal to the number of vertices ($n$) in the graph. However, sometimes there are vertices where local complementation does not have any effect. Moreover, there could also be vertices that lead to isomorphic graphs after applying local complementation, as one could observe in the work of Adcock et al.~\cite{adcock2020mapping}. If the neighbouring states are chosen at random, these vertices could affect the performance of the SA algorithm. Therefore, we use a graph metric called the local clustering coefficient ~\cite{clustering_coeff} to guide the SA algorithm to decide where local complementation should be applied, thereby improving the efficiency of SA algorithm.

The local clustering coefficient $C_G(v)$ of a vertex of a graph (which will always be clear from the context) provides information on the connectedness of its neighbourhood. Specifically, it is the ratio between the number of edges connecting the neighbours of the given vertex and the total number of possible edges. This metric has a maximum value of 1 (when the neighbourhood is fully connected) and minimum value of zero (when the neighbours share no other edges). That is, 
\begin{align}
    C_G(v) = \frac{\text{Number of edges between neighbours}}{\text{Possible connections between neighbours}}.
\end{align}
A local complementation ($L_v$) on a vertex $v$ affects its local clustering coefficient $C_G(v)$ in the following way,
\begin{equation}
    C_{L_v(G)}(v) = 1 - C_G(v).
\end{equation}
Therefore, applying a local complementation to a vertex $v$ with $C_G(v)\geq \frac{1}{2}$ (i.e.~a more connected neighbourhood) changes the local clustering coefficient to be smaller than $\frac{1}{2}$ (i.e.~a less connected neighbourhood). Thus, applying local complementations to vertices with a high value of $C_G(v)$ reduces the local edge density. Moreover, to bias vertices that have a larger number of neighbours, we multiply $C_G(v)$ with the degree of the vertex, leading to the heuristic quantity $M_v\equiv C_G(v)D_v$, similar to~\cite{pgs_lc}. Based on this heuristic, we decide on the set of states $S(s_k)$ from which the SA algorithm could choose the potential solution $s_\textrm{pot}$. Thus, we bias SA into choosing vertices that have a higher value of $M_v$.

\begin{figure}[t]
    \centerfloat
    \begin{minipage}[t]{0.49\textwidth} 
        \centerfloat\includegraphics[width=0.9\textwidth]{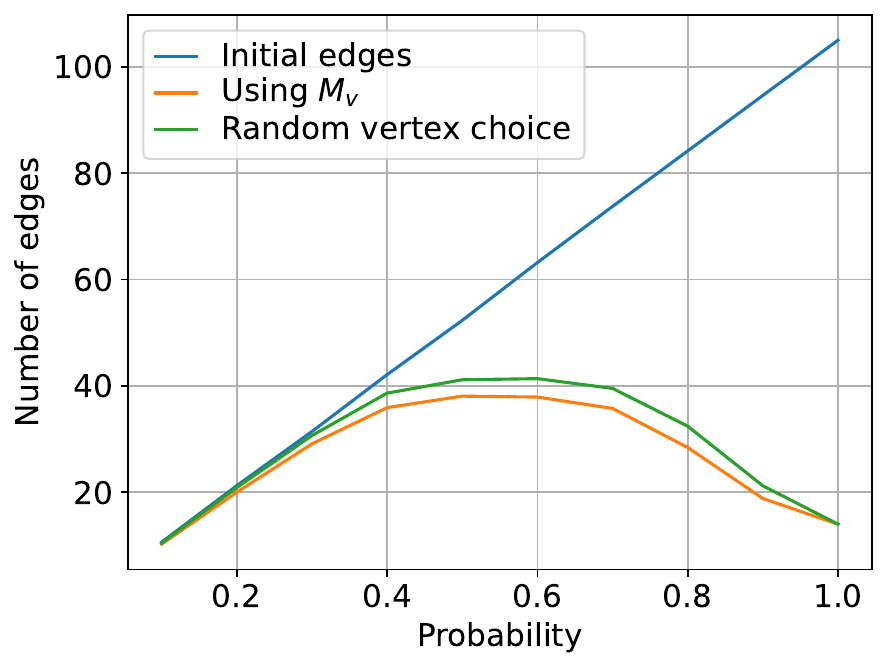}
    \end{minipage}
    \caption{Here we plot the average edges in the initial graphs (blue line) and the edges obtained from EDM-SA using metric $M_v$ (orange) and random vertex choice (green). In the random vertex choice method, a potential solution $s_\textrm{pot}$ is found by applying local complementation to a vertex chosen uniformly at random. Initial edges are the edges in the input ER graph. For this run, $k_{\textrm{max}}$ and $T(1)$ were both set to 100.}
    \label{fig: sampling}
\end{figure}

\begin{figure*}[t]
    \centerfloat
    \begin{minipage}[t]{0.99\textwidth} 
        \centerfloat\includegraphics[width=0.99\textwidth]{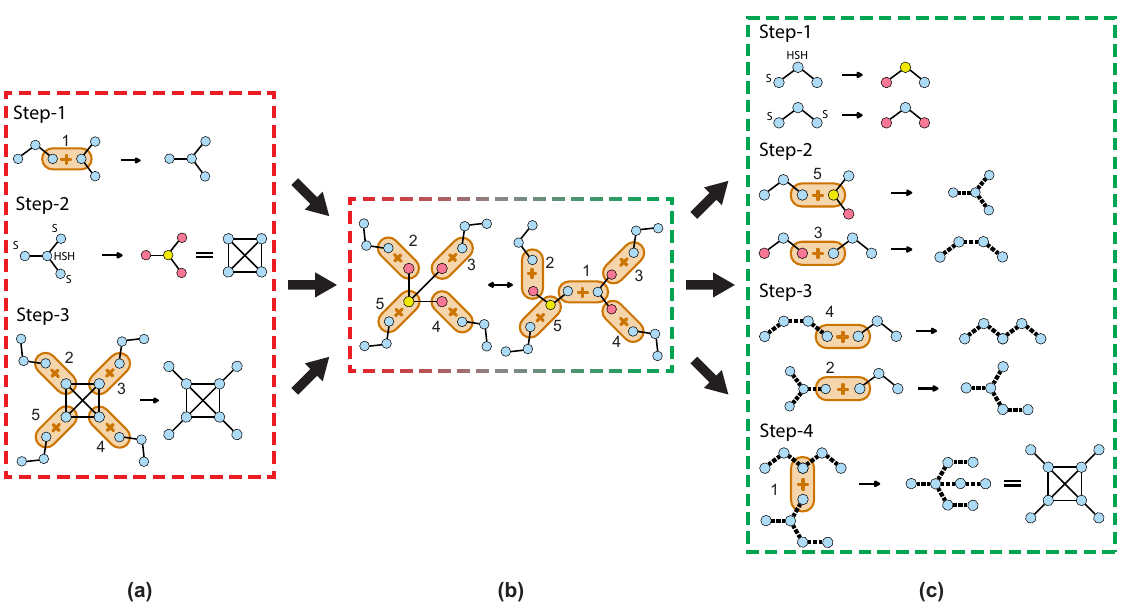}
    \end{minipage}
    \caption{\textbf{(a)} shows the order of fusions for constructing an RGS with 8 qubits following the Intermediate-LC protocol. This order can be used to derive the fusion network shown in \textbf{(b)}. The fusion network enables the tracking of the LC-gates relative to fusions for each resource state. {To track each fusion operation, we label each fusion with a number next to it. These numbers allow us to keep track of the fusion operations after we commute the LC-operations through the fusions.} Applying LC-gates to resource states in the beginning, as shown in the fusion network, allows the fusions to be reordered flexibly. This reordering can be optimised to reduce the resource requirements. An example of such reordering is the creation of the graph $H$+$L$, as shown in \textbf{(c)}.}
    \label{fig: grgs_commutation}
\end{figure*}

The algorithm finds the set $S(s_k)$ at the $k$'th iteration based on the values of heuristic $M_v$, $k$ and $k_{\textrm{max}}$. We first construct a set with values of $M_v$ of each vertex in the state $s_k$. We then find a subset of this set with unique values of $M_v$. Then, we sample a value from the set of these unique values based on a cutoff. The cutoff is given by $c = k/k_{\textrm{max}}\cross l$, where $l$ is the number of elements in the set of unique values. We randomly select a value from the set of unique values that is greater than the value corresponding to the cutoff; let us call the selected value $M_u$. Since $M_u$ is in the set of unique values, multiple vertices could have their metrics equal to $M_u$. Therefore, we find all the vertices that have $M_u$ as their metric value. The graphs obtained by applying local complementation once to these vertices then constitute the set $S(s_k)$. We then sample a state from this set uniformly at random to find the next candidate state $s_{\textrm{pot}}$.% In conclusion, we bias the selection process to choose vertices with higher values of $M_v$ and apply local complementation to that vertex to fin.

%We first find a set ($T(s_K) = \{M_1, M_2,\ldots, M_n\}$) that contains the value of $M_v$ for $v \in \{1,\ldots,n\}$ of graph $G_k$. We find the set of unique values from this set arranged in non-decreasing order, called the $U^{(k)} = \{u^{(k)}_1, u^{(k)}_2, \ldots, u^{(k)}_l\}$. At the $k$'th iteration, we find a cutoff $c = k/k_{\textrm{max}}\cross l$, where $l$ is number of elements in the set. Then a value $u_i^{(k)}$ is chosen uniformly at random from this set such that $i \geq c$. We find the vertices in $T(s_K)$ that have their metric value $M_v$ equal to the sampled value $u_i^{(k)}$. 

We benchmark this procedure of sampling vertices using the metric $M_v$ against a procedure where vertices are chosen uniformly at random. We run simulated annealing for ER graphs $G(n,p)$ with 15 vertices for increasing probability $p$ in Fig.~\ref{fig: sampling}. We generate 100 graphs for values of $p = \{0.1, 0.2, 0,3,\ldots, 1 \}$ and we apply the SA procedure to each graph. We find that by using $M_v$ as our guiding metric, SA can find better solutions compared to sampling at random.

\section{Generalised repeater graph states}\label{sec: grgs_app}

Here, we discuss our method for creating all-photonic generalised repeater graph states (gRGS). We consider using single photon sources and optical fusions~\cite{fusion} to create the desired graph state. To find the optimal number of resources required for creating gRGS, we use the protocol from Ref.~\cite{Lee_2023}. We combine our algorithms with their work to find an improvement in the number of required fusions and the number of resource states (3-qubit GHZ states) for efficiently creating gRGS. 

A gRGS has an $n$-vertex central graph $G$ with leaves $L$ connected to each vertex, as shown in Fig.~\ref{fig: grgs_construction}(b)---we will refer to this graph as $G$+$L$. Using the EDM-SA algorithm, we can reduce the number of edges in the graph $G$ by finding an approximate MER $H$. Thus, one way to create a gRGS is the Intermediate-LC protocol where we first create the graph $H$, followed by the application of LC-gates to convert $H$ to $G$. The leaves are then fused to the central graph $G$ to create $G$+$L$.

The above approach is suboptimal, however. This is because the order in which fusions are performed is restricted to first performing the fusions that generate the graph $H$, while the fusions that involve attaching the leaves need to happen at the end. Attaching individual leaves at the end, one by one, to the central graph $G$ will lead to an exponentially decaying success probability with the number of leaves due to the probabilistic nature of fusions. In general, probabilistic fusions of two graphs of very different sizes are considered suboptimal.

To find a better fusion order, we propose applying LC-gates corresponding to the $H$ to $G$ conversion to the resource states \emph{before} beginning the fusion process. This leads to the following steps: (1) apply LC-gates, (2) follow fusion order to create $H$, and (3) fuse leaves $L$. We observe that fusions commute with each other as they act on different qubits. Therefore, we can combine the construction of $H$ and the addition of leaves $L$ into a single step. This simplifies our protocol into two steps: (1) apply LC-gates to suitable qubits of the resource states, (2) follow the fusion order for creating $H$+$L$ using these modified resource states. The final state from this fusion order is $G$+$L$ due to the LC-gates applied at the beginning.

Commuting the LC-gates to the beginning of the protocol can be understood by stepping away from the graph state picture and considering the circuit of the protocol. Consider the steps in construction of $G$+$L$ following the Intermediate-LC protocol: (1) construct the graph $H$, (2) apply LC-gates to convert $H$ to $G$, and (3) fuse leaves $L$ to the graph $G$. We make the following observations from the circuit picture: (1) the target qubits for LC-gates have not been involved in any fusions until that point in time (otherwise they would have been measured out). As a result, the LC-gates can be pushed to the beginning of the protocol as they commute with the fusions implemented to construct $H$. (2) Pauli corrections may be required on qubits adjacent to those involved in fusions. If an LC-gate acts on such a qubit, the corrections can be modified accordingly. This is possible since the Clifford group normalises the Pauli group, which allows us to commute the LC-gates through the Pauli corrections. Notably, Ref.~\cite{Lee_2023} also applies LC-gates corresponding to local complementations. These gates must also be accounted for when commuting LC-gates to the beginning.

In Fig.~\ref{fig: grgs_commutation}, we demonstrate the construction of an RGS by commuting LC-gates to the beginning. We first find the suboptimal fusion order for creating $G$+$L$ using Intermediate-LC, shown in Fig.~\ref{fig: grgs_commutation}(a). We then construct the \emph{fusion network} (similar to Ref.~\cite{Lee_2023}) to determine the location of each LC-gate. A fusion network depicts the order in which each resource state has to be fused. We show the fusion network corresponding to Intermediate-LC in Fig.~\ref{fig: grgs_commutation}(b). The fusion network also details the information about the LC-gates in the context of the required fusions. We use this information to apply the LC-gates to the appropriate qubits of the resource states at the start. This allows us to reorder the fusions to optimise the number of resources.

The LC-gates corresponding to a local complementation operation on a node $v$ correspond to the application of the HSH gate on the qubit corresponding to node $v$ and the S gate to all the qubits that correspond to the neighbours of node $v$. These gates altogether convert a graph state into another graph state. Applying a subset of these gates to the graph state will result in a state that is not a graph state but is LC-equivalent to a graph state: a \emph{non-graph state}. When we commute LC-gates corresponding to the conversion of graph $H$ to $G$, we distribute these gates among multiple resource states, as shown in Fig.~\ref{fig: grgs_commutation}(c). Thus, the resource states become non-graph states (but are still LC-equivalent) after the application of these LC-gates. We then fuse these non-graph states following the fusion order for creating $H$+$L$ to obtain our desired graph state of $G$+$L$. Thus, the final equality in Step-4 of Fig.~\ref{fig: grgs_commutation}(c) takes a non-graph state (LHS) to a graph state (RHS).

In conclusion, the Commute-LC protocol enables the creation of $G$+$L$ by following the procedure for constructing $H$+$L$. Importantly, applying LC-gates to the resource states converts them into LC-equivalent non-graph states. The locations of the LC-gates can be identified from the fusion network for creating $G$+$L$ via Intermediate-LC. This approach has two main advantages: (1) it reduces the number of required fusions and resource states by creating $H$+$L$, and (2) it enables obtaining the optimal fusion order using \texttt{OptGraphState} from Ref.~\cite{Lee_2023}.

\section{{Example of MERs}}

In this section we illustrate the difference between the edge minimisation performance of EDM-SA and EDM-SAILP. 
\begin{figure*}[h]
    \centerfloat
    \begin{minipage}[t]{0.99\textwidth} 
        \centerfloat\includegraphics[width=0.99\textwidth]{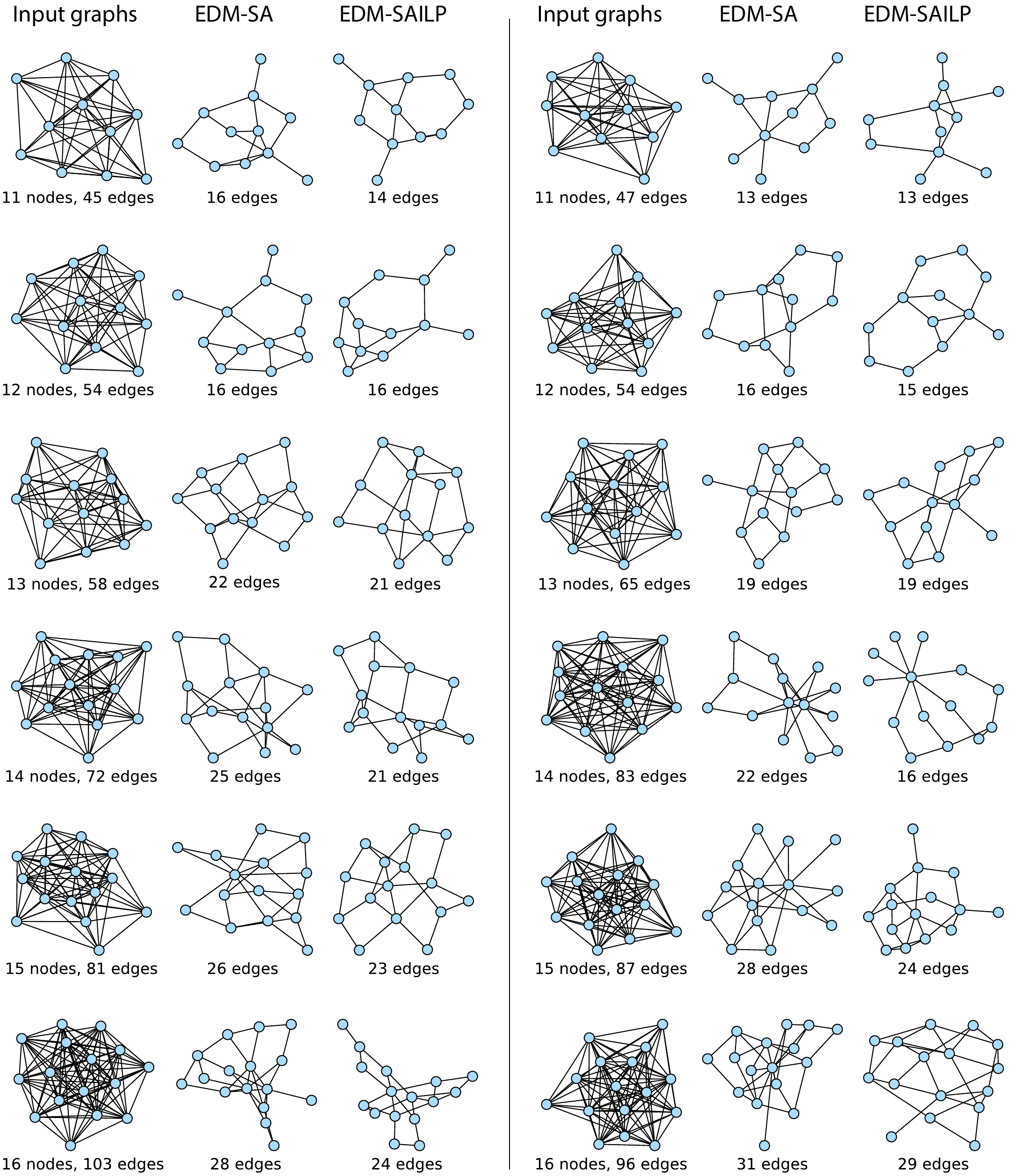}
    \end{minipage}
    \caption{We sample ER graphs $G(n,p)$ with increasing $n$ from 11 to 16 and $p=0.8$. We run EDM-SA and EDM-SAILP on these graphs, with $k_{\textrm{max}}$ and $T(1)$ set to 100. We illustrate a set of three graphs, namely: (1) input graphs, (2) approximate MERs obtained from EDM-SA and (3) the exact MERs obtained from EDM-SAILP. Each row has two such sets with the same number of vertices, and the number of vertices increase from 11 to 16 over all rows. We can observe that both the algorithms provide significant reductions in the number of edges, this is partly be due to the input graph being dense. We also observe that in some cases the number of edges in the approximate MERs obtained from EDM-SA are a little higher than the MERs.}
    \label{fig: mer_examples}
\end{figure*}

\end{document}

%% file: lcexample.tex
\begin{tikzpicture}[
  parse node name/.style={
    circle, minimum size=3.5mm, draw,
    fill={rgb,255:red,171; green,219; blue,248}
  },
  decide color/.style 2 args={
    /utils/TeX/if=c#1
      {/utils/TeX/ifnum={#2<5}{bluelight}{bluedark}}
      {/utils/TeX/ifnum={#1<8}{light}{dark}}
  },
  light/.style={fill=gray!20},  
  bluelight/.style={fill=blue!10},
  dark/.style={fill=gray!60},  
  bluedark/.style={fill=blue!30}
]

% Second figure (originally on the right) now moved to the left
\def\ra{0}

% \node[parse node name] (1) at ({sin(0*360/5)+\ra}, {-cos(0*360/5)}) {};
% \node[parse node name] (3) at ({sin(2*360/5)+\ra}, {-cos(2*360/5)}) {};
% \node[parse node name] (4) at ({sin(3*360/5)+\ra}, {-cos(3*360/5)}) {};
% \node[parse node name, scale=1] (0) at (0.0+\ra, -0) {};

\draw[line width=0.2mm] ({sin(2*360/5)+\ra}, {-cos(2*360/5)}) -- ({0+\ra}, {-1});
\draw[line width=0.2mm] ({sin(3*360/5)+\ra}, {-cos(3*360/5)}) -- ({0+\ra}, {-1});
\draw[line width=0.2mm] (0+\ra,-1) -- (0.8+\ra, -1.2);
\draw[line width=0.2mm] (0+\ra,-1) -- (-0.7+\ra, -1.3);
\draw[line width=0.2mm] ({sin(2*360/5)+\ra}, {-cos(2*360/5)}) -- (1.1+\ra, 0.6);
\draw[line width=0.2mm] ({sin(2*360/5)+\ra}, {-cos(2*360/5)}) -- (1.2+\ra, 1.3);
\draw[line width=0.2mm] ({sin(2*360/5)+\ra}, {-cos(2*360/5)}) -- (0.4+\ra, 1.6);
\draw[line width=0.2mm] ({sin(3*360/5)+\ra}, {-cos(3*360/5)}) -- (-1.1+\ra, 0.4);
\draw[line width=0.2mm] ({sin(3*360/5)+\ra}, {-cos(3*360/5)}) -- (-1.2+\ra, 1.1);
\draw[line width=0.2mm] (0+\ra,-1) -- (0+\ra, -0);
\draw[line width=0.2mm] ({sin(2*360/5)+\ra}, {-cos(2*360/5)}) -- (0+\ra, -0);
\draw[line width=0.2mm] ({sin(3*360/5)+\ra}, {-cos(3*360/5)}) -- (0+\ra, -0);

% Arrow
\draw [-stealth](1.8, 0) -- (2.5, 0);

% First figure (originally on the left) now moved to the right
\def\ra{4.2}

\node[parse node name] (1) at ({sin(0*360/5)}, {-cos(0*360/5)}) {};
\node[parse node name] (3) at ({sin(2*360/5)}, {-cos(2*360/5)}) {};
\node[parse node name] (4) at ({sin(3*360/5)}, {-cos(3*360/5)}) {};
\node[parse node name, scale=1,fill={rgb,255:red,248; green,119; blue,148}] (0) at (0.0, -0) {};

\draw[line width=0.2mm] ({sin(2*360/5)+\ra}, {-cos(2*360/5)}) -- ({sin(3*360/5)+\ra}, {-cos(3*360/5)});
\draw[line width=0.2mm] (0+\ra,-1) -- (0.8+\ra, -1.2);
\draw[line width=0.2mm] (0+\ra,-1) -- (-0.7+\ra, -1.3);
\draw[line width=0.2mm] ({sin(2*360/5)+\ra}, {-cos(2*360/5)}) -- (1.1+\ra, 0.6);
\draw[line width=0.2mm] ({sin(2*360/5)+\ra}, {-cos(2*360/5)}) -- (1.2+\ra, 1.3);
\draw[line width=0.2mm] ({sin(2*360/5)+\ra}, {-cos(2*360/5)}) -- (0.4+\ra, 1.6);
\draw[line width=0.2mm] ({sin(3*360/5)+\ra}, {-cos(3*360/5)}) -- (-1.1+\ra, 0.4);
\draw[line width=0.2mm] ({sin(3*360/5)+\ra}, {-cos(3*360/5)}) -- (-1.2+\ra, 1.1);
\draw[line width=0.2mm] (0+\ra,-1) -- (0+\ra, -0);
\draw[line width=0.2mm] ({sin(2*360/5)+\ra}, {-cos(2*360/5)}) -- (0+\ra, -0);
\draw[line width=0.2mm] ({sin(3*360/5)+\ra}, {-cos(3*360/5)}) -- (0+\ra, -0);

\node[parse node name] (5) at ({sin(0*360/5)+\ra}, {-cos(0*360/5)}) {};
\node[parse node name] (6) at ({sin(2*360/5)+\ra}, {-cos(2*360/5)}) {};
\node[parse node name] (7) at ({sin(3*360/5)+\ra}, {-cos(3*360/5)}) {};
\node[parse node name, scale=1,fill={rgb,255:red,248; green,119; blue,148}] (8) at (0.0+\ra, -0) {};

\end{tikzpicture}